\documentclass[12pt,oneside, a4paper]{article}
\pdfoutput=1

\ifx\pdfoutput\undefined
\usepackage[dvips,bookmarks=false]{hyperref}	This is for arXiv.org
\else
\usepackage{hyperref}	
\fi
\hypersetup{colorlinks,bookmarksopen,bookmarksnumbered,citecolor=blue,
linkcolor=blue,pdfstartview=FitH,urlcolor=blue}

\usepackage{graphicx}
\usepackage{amssymb}
\usepackage{cite}
\usepackage{bm}
\usepackage{amsmath,amsthm}

\newcommand{\capdef}{}
\newcommand{\mycaption}[2][\capdef]{\renewcommand{\capdef}{#2}%
       \caption[#1]{{\footnotesize #2}}}
\newcommand{\be}{\begin{equation}}
\newcommand{\ee}{\end{equation}}
\newcommand{\vel}{{\nu}}

\begin{document}


\begin{center}

\vspace{1cm}
{\Large\bf Higgs portal, fermionic dark matter, and\\[2mm] 
a Standard Model like Higgs at 125 GeV}
\vspace{1cm}

\renewcommand{\thefootnote}{\fnsymbol{footnote}}
{\bf Laura Lopez-Honorez}$^{1,}$\footnote[1]{llopezho AT mpi-hd.mpg.de},
{\bf Thomas Schwetz}$^{1,}$\footnote[2]{schwetz AT mpi-hd.mpg.de},
{\bf Jure Zupan}$^{2,}$\footnote[3]{jure.zupan AT cern.ch}
\vspace{5mm}

{\it%
$^{1}${Max-Planck-Institut f\"ur Kernphysik, Saupfercheckweg 1, 69117 Heidelberg, Germany}\\
$^2$Department of Physics, University of Cincinnati, Cincinnati, Ohio 45221,USA
}

\vspace{8mm} 

\abstract{We show that fermionic dark matter (DM) which communicates with
the Standard Model (SM) via the Higgs portal is a viable scenario, even if a
SM-like Higgs is found at around 125~GeV. Using effective field theory we
show that for DM with a mass in the range from about 60 GeV to 2 TeV the
Higgs portal needs to be parity violating in order to be in agreement with
direct detection searches. For parity conserving interactions we identify
two distinct options that remain viable: a resonant Higgs portal, and an
indirect Higgs portal. We illustrate both possibilities using a simple
renormalizable toy model.}

\end{center}

\renewcommand{\thefootnote}{\arabic{footnote}}
\setcounter{footnote}{0}


\section{Introduction} 

How dark matter (DM) couples to Standard Model (SM) particles is an open
question. An interesting possibility is that the coupling of dark and
visible sectors is through a Higgs portal \cite{Patt:2006fw, Kim:2006af,
MarchRussell:2008yu, Kim:2008pp, Ahlers:2008qc, Feng:2008mu, Andreas:2008xy,
Barger:2008jx, Kadastik:2009ca, Kanemura:2010sh, Piazza:2010ye,
Arina:2010an, Low:2011kp, Djouadi:2011aa, Englert:2011yb, Kamenik:2012hn,
Gonderinger:2012rd, Lebedev:2012zw}. The operator $(H^\dagger H)$ is one of
the two lowest dimensional gauge invariant operators that one can write in
the SM (the other one being the hyper-charge gauge field strength
$B_{\mu\nu}$). Therefore, it is quite likely that also $(H^\dagger
H)$--(dark sector) will be the lowest dimension operator connecting dark and
visible sectors, and thus potentially the most important one.

Experimentally the Higgs portal is probed from two complementary
directions.  On the one hand, the new generation of direct DM
detection experiments \cite{Aprile:2011hi, Ahmed:2011gh} is starting
to probe DM--nucleon scattering cross sections of roughly the size
given by a single Higgs exchange with the SM Yukawa couplings to the
quarks. On the other hand, the first hints of a SM-like Higgs boson
signal were reported by the ATLAS and CMS collaborations. The hints of
a signal are seen in several channels, pointing to a Higgs mass of
roughly $m_h\sim 124-126$ GeV \cite{:2012si,Chatrchyan:2012tx}, with
the SM Higgs boson consistent with the current data at 82\%
C.L.~\cite{Espinosa:2012ir}, see also~\cite{Azatov:2012bz,
  Carmi:2012yp}.  Those hints are supported by recent results from D0
and CDF. For updates from ATLAS, CMS, D0, and CDF see \cite{moriond}.

In view of these experimental developments we revisit the Higgs portal to
DM. In particular we focus on fermionic DM. The Higgs
portals for the fermionic DM and the scalar DM are qualitatively different. 
For instance, if DM is a scalar, $\phi_{DM}$, then the Higgs portal operator
$(H^\dagger H) (\phi_{DM}^\dagger \phi_{DM})$ is renormalizable. The same is
true if DM is a spin-1 particle. In contrast, if DM is a fermion, $\chi$,
then the Higgs portal necessarily proceeds through non-renormalizable
interactions. The lowest dimensional Higgs portal in that case consists of
two dim~$=5$ operators
\begin{equation}\label{Qs}
Q_1=(H^\dagger H) (\bar \chi \chi) \,, \qquad Q_5=i(H^\dagger H) (\bar \chi
\gamma_5 \chi) \,,
\end{equation}
which enter the effective Hamiltonian
\begin{equation} \label{eq:Heff}
H_{\rm eff}=\frac{1}{\Lambda_1}Q_1+\frac{1}{\Lambda_5}Q_5 \,.
\end{equation}
The mass scales $\Lambda_{1,5}$ are roughly the masses of the mediators for
${\mathcal O}(1)$ couplings between DM and the mediators. Since the DM--Higgs
effective Hamiltonian is non-renormalizable, this means that a Higgs portal
for fermionic dark matter necessarily requires a UV completion. In this
paper we also consider situations when such UV completions are required 
in order to obtain a correct description of the DM phenomenology.

We first use the effective field theory (EFT) description of the Higgs
portal \eqref{eq:Heff} and derive consequences for each of the two operators
$Q_{1,5}$.  The parity conserving interaction, $Q_1$, is severely
constrained by direct detection experiments. If only $Q_1$ is present in
$H_{\rm eff}$ then one cannot obtain a small enough relic density consistent
with the bound from XENON100 for DM masses below about 2~TeV
\cite{Djouadi:2011aa}. In contrast, as we will show in
section~\ref{sec:EFT}, the parity violating operator $Q_5$ is allowed by
direct detection searches and the observed relic density can be obtained,
see also \cite{Pospelov:2011yp}. Hence, when DM interactions are mediated by
fields much heavier than $2m_{\chi}$ and $2m_h$ the EFT description is valid
and we must conclude that the Higgs portal interactions for fermionic DM
need to be parity violating (``pseudo-scalar Higgs portal''). Yet viable
scenarios with parity conserving operators can be found when EFT breaks
down. We identify two distinct options:
\begin{itemize}
\item
``resonant Higgs portal" -- where the dominant contribution is due to a
resonant annihilation either through the Higgs or the mediator, 
\item
``indirect Higgs portal" -- where the DM annihilations into the mediator set
the relic density and the Higgs portal only provides the link between the visible
and dark sector thermal baths.
\end{itemize}
In section~\ref{sec:model} we illustrate both of these two possibilities
using a toy model -- the minimal extension of the SM with a DM fermion
$\chi$ and a real singlet $\phi$ see e.g.~\cite{Kim:2008pp, Baek:2011aa}.

\section{Effective field theory considerations}
\label{sec:EFT}

Let us first assume that the mediators are heavy so that they can be
integrated out. The Higgs portal is then given by
eq.~\eqref{eq:Heff}. We will be interested in the direct detection
of DM and in the annihilation of DM in the galactic halo. In both cases the
DM particles entering the process are non-relativistic, with velocities
typical of DM in the galactic halo, $\vel \sim 10^{-3}$. For annihilations
in the early universe, responsible for obtaining the thermal relic density,
DM is moderately relativistic.

The indirect and direct DM detection signals are given by the annihilation
of two non-relativistic DM particles and the scattering of non-relativistic DM
on the nuclei, respectively. For these two processes the two effective
operators $Q_{1,5}$ behave in exactly the opposite way.  For instance, the
annihilation cross section is (for a Majorana fermion $\chi$)
\begin{equation}
\sigma_{\rm ann}=\frac{1}{4\pi}\left[\frac{\left(1-4 m_\chi^2/s\right)}{\Lambda_1^2}+\frac{1}{\Lambda_5^2}\right] \frac{f(m_\chi)}{\sqrt{1-4m_\chi^2/s}},
\end{equation}
which is in the non-relativistic limit
\begin{equation}\label{eq:ann-nonrel}
\sigma_{\rm ann}=\frac{1}{4\pi\vel}\left[\frac{\vel^2}{\Lambda_1^2}+\frac{1}{\Lambda_5^2}\right] f(m_\chi),
\end{equation}
where $\vel$ is the velocity of each of the DM particles in the
center-of-mass system (CMS). The contributions to annihilations from the
parity conserving operator $Q_1$ are thus velocity suppressed, while parity
violating contributions, due to $Q_5$, are unsuppressed. The function
$f(m_\chi)\equiv \sum_i f_i$ sums the available final states $i$. For
instance, for $m_\chi>m_h$ we have
\begin{equation}
f_h=\left(1+\frac{3 m_h^2}{s- m_h^2}\right)^2 \left(1-4m_h^2/s\right)^{1/2}, \qquad f_t=\frac{m_t^2}{s}\frac{(1-4m_t^2/s)^{3/2}}{(1-m_h^2/s)^2},
\end{equation}
for the decays to Higgs and top, respectively. For $m_\chi$ very heavy
$f_h\to 1$ and $f_t\to 0$. 

In the
early universe, around the freeze-out temperature $T_F$ we have 
$\vel^2 \sim T_F / m_\chi \simeq 1/20$, whereas in the galactic halo we have
$\vel^2 \sim 10^{-6}$. As a consequence, for parity conserving interactions
the annihilation cross section relevant for indirect detection signals is
significantly suppressed compared to the one relevant for 
thermal freeze-out. In contrast, for parity violating interactions the
annihilation cross section is independent of the velocity.

For direct detection the situation is exactly opposite. 
Integrating out the Higgs field, the scattering of DM on matter is given by
an effective Hamiltonian
\begin{equation}
H=\sum_{i=1,5}
\frac{C_i}{m_h^2} O_i, 
\end{equation}
where the two operators and Wilson coefficients are
\begin{equation}
O_{1,5}=\bar \chi \{1,i \gamma_5\}\chi \sum_q \frac{m_q}{v}\bar q q,\qquad C_{1,5}=\frac{v}{\Lambda_{1,5}}.
\end{equation}
where $v=246$  GeV is the Higgs vacuum expectation value (VEV). 
The cross section for $\chi$ scattering on the proton induced by the operator $O_1$ 
is then 
\begin{equation}\label{eq:scattering}
\sigma(\chi p\to \chi p)=\frac{4}{\pi}\left(C_1 g_{Hp} \right)^2 \left(\frac{m_{\rm red}}{m_h^2}\right)^2,
\end{equation}
where $m_{\rm red} = m_p m_\chi / (m_p + m_\chi)$ is the reduced mass of the
DM--proton system and
\begin{equation}
 g_{Hp}=\frac{m_{p}}{v}\left[\sum_{q=u,d,s}f^{(p)}_{q}+\frac{2}{9}\left(1-
\sum_{q=u,d,s}f^{(p)}_{q}\right)\right] \approx 1.3\times 10^{-3}\,,
\end{equation}
see e.g., \cite{Belanger:2008sj}, where also values for $f^{(p)}_q$ are
given. 
The operator $O_5$, on the other hand, induces a velocity suppressed scattering
\begin{equation}
\sigma(\chi p\to \chi p)=\frac{2}{\pi}\left(C_5 g_{Hp} \right)^2 \left(\frac{m_{\rm red}}{m_h^2}\right)^2 \vel^2.
\end{equation}
where typically $\vel \sim 10^{-3}$. Hence in direct detection one obtains a
velocity suppressed scattering cross section for parity violating
interactions, but unsuppressed scattering for parity conserving ones. 

\begin{figure}[tb!]
\begin{center}
\includegraphics[width=10cm]{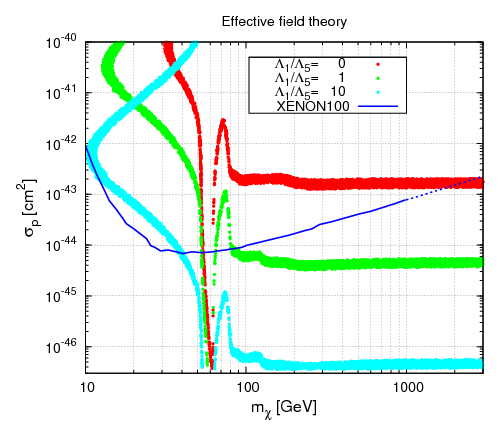} 
 \mycaption{Proton--dark matter scattering cross
 section as a function of the dark matter mass in the
 effective field theory of eq.~(\ref{eq:Heff}), as predicted by 
 requiring that the correct relic density is obtained by thermal freeze-out.
 The scattering cross section is shown for several ratios
 of pseudo scalar coupling to scalar coupling $\Lambda_1/\Lambda_5$, and
 compared to the limit from XENON100~\cite{Aprile:2011hi}. }
 \label{fig:DDEFT}
\end{center}
\end{figure}

This means that it is possible to find regions of $C_{1,5}$ parameter
space with correct relic density but small enough direct detection
signals. In fig.~\ref{fig:DDEFT} we show the scattering cross section
for different ratios of pseudo scalar coupling to scalar coupling
$\Lambda_1/\Lambda_5$.  For a fixed ratio and given DM mass we
determine the sizes of $\Lambda_i$ by requiring that the thermal relic
density is obtained in the interval $0.09<\Omega h^2<0.13$. For the
thermal freeze-out calculations we use the {\tt
  micrOMEGAs}~\cite{Belanger:2008sj, Belanger:2010gh} public code, and
we assume $m_h = 125$~GeV. The curve for $\Lambda_1/\Lambda_5 = 0$
corresponds to parity conservation, i.e., pure scalar coupling. This
case was considered for example in \cite{Kim:2006af, Kanemura:2010sh,
  Djouadi:2011aa} and is incompatible with the bound from
XENON100~\cite{Aprile:2011hi} for DM masses $m_\chi \lesssim
2$~TeV. However, already for $\Lambda_1/\Lambda_5 \simeq 1$ the
predicted cross section is below the XENON100 bound for $m_\chi
\gtrsim 100$~GeV, while $\Lambda_1/\Lambda_5 \simeq 10$ leads to cross
sections of about two to three orders of magnitudes below the limit.

Around $m_\chi \approx m_h/2 = 63$~GeV the effect of the $s$-channel
resonance due to Higgs exchange is clearly visible in fig.~\ref{fig:DDEFT}.
For DM masses below the resonance Higgs decays $h\to 2\chi$ become possible.
The shape of the curves below the resonance is due to the Breit-Wigner form
of the annihilation cross section. The contribution of $h\to 2\chi$ to the
Higgs decay width allows for two solutions for $\Lambda_i$ giving rise to
the correct relic density at a given DM mass above a certain minimal mass
and below the resonance. Note however, that in some cases $\Lambda_i$ may
become even smaller than $m_\chi$ and the EFT description may no longer be
valid. Moreover, typically large branching fractions of the Higgs into DM
are obtained in those cases. Therefore, the region below $m_h/2$ would be
excluded by observing a Higgs with SM-like decay branching fractions.

Let us briefly mention constraints from indirect detection. In the case of
pure scalar interactions DM annihilations are $\vel^2$ suppressed. In the
early universe at freeze-out $\vel^2 \simeq 1/20$, whereas in the galactic halo we have
$\vel^2 \sim 10^{-6}$. This leads to a negligible signal for indirect
detection experiments. As soon as the pseudo-scalar coupling becomes
comparable to the scalar one, the annihilation cross section is dominated by
pseudo-scalar interactions with $\sigma \vel$ independent of the velocity,
see eq.~\eqref{eq:ann-nonrel}. Therefore, in the latter case, the annihilation cross section
will be the ``thermal'' one with $\sigma \vel \simeq 3\times 10^{-26}\,\rm
cm^3s^{-1}$. Current data from FERMI-LAT and BESS-Polar~II disfavour such
cross sections for $m_\chi \lesssim 30$~GeV \cite{Ackermann:2011wa,
Kappl:2011jw}. Since here we are restricted to DM masses $m_\chi \gtrsim
60$~GeV current limits from indirect detection do not constrain this model.

Let us mention that monojet searches for dark matter at colliders~\cite{Beltran:2010ww,Bai:2010hh} will apriori not constrain further this EFT model of dark matter. Limits on spin-independent interactions from recent LHC data~\cite{Fox:2011pm,cms} are much weaker than Xenon bounds in the region of interest.  

\section{Beyond the EFT framework}
\label{sec:model}

\subsection{The toy model}

Now we move to situations which cannot be described by the EFT. In
order to illustrate when it is possible to have a viable fermionic DM Higgs
portal we consider a simple UV completion by introducing a real
scalar singlet that will act as mediator particle\footnote{For alternative UV completions see e.g.~\cite{Bird:2006jd,D'Eramo:2007ga}.}. For simplicity we consider a  
Majorana fermion $\chi$ as DM, with $\chi = \chi_L + \chi_L^c$ in
4-component notation. (All our arguments will equally apply to the
Dirac case.) We denote the SM Higgs doublet by $H$ and the real
singlet scalar by $\varphi$. The relevant terms in the Lagrangian are
\begin{align}
\mathcal{L} &= 
\frac12 \bar\chi_L(i \gamma_\mu\partial^\mu - \mu_\chi - g \varphi)  \chi_L^c +
  {\rm h.c.} \nonumber\\
&+
(D_\mu H)^\dagger D^\mu H +
\frac{1}{2} \partial_\mu\varphi\partial^\mu\varphi 
- V(\varphi, H)\,,
\end{align}
Here $D^\mu$ is the SM gauge-covariant derivative, $V(\varphi, H)$ is the
Higgs potential, and we allow the coupling constant $g$ and the mass
parameter $\mu_\chi$ to be complex. Let us work in 
unitary gauge and expand $H$ and $\varphi$ around their VEVs:
\begin{equation}
H= \frac{1}{\sqrt{2}} \left(\begin{array}{c} 0 \\ h+v_1 \end{array}\right)
\,,\qquad \varphi= \phi+v_2
\end{equation}
where $v_1 = 246$~GeV. By performing a phase transformation 
$\chi_L \to e^{i\alpha/2}
\chi_L$ with $\alpha = \mathrm{Arg}(\mu_\chi + g v_2)$ we find that the
physical mass of $\chi$ corresponds to $m_\chi = |\mu_\chi + g v_2|$.
The phase of $g$ relative to the mass term determines the scalar ($S$) or
pseudo-scalar ($P$) nature of the Yukawa coupling:
\begin{equation}
g_S = \mathrm{Re}(g e^{-i\alpha}) \,,\qquad
g_P = \mathrm{Im}(g e^{-i\alpha}) \,.
\end{equation}
A non-zero value of $g_P$ violates parity. 
The mass term and the
interaction terms for the Majorana fermion $\chi$ become thus:
\begin{equation}
  {\cal L}_\chi=-\frac12 \left(m_\chi \bar \chi \chi+ g_S \phi \bar \chi
  \chi +i g_P \phi \bar \chi \gamma_5 \chi \right) \,.
\end{equation}
As discussed in the previous section the pseudo-scalar coupling leads
to suppressed rates in direct detection. Therefore, it is always
possible to consider the situation of $g_S \ll g_P$ in order to
reconcile the annihilation cross section required for the relic
density with stringent bounds on the DM--nucleon scattering cross
section. In the following we discuss alternative ways to
achieve this goal, and therefore we assume in this section parity
conservation, $g_P = 0$, keeping always in mind the possibility of
parity violation on top of the mechanisms discussed here.

The Higgs potential is
\begin{align}
V(\varphi, H) =& -\mu_H^2 H^\dagger H + \lambda_H (H^\dagger H)^2
              - \frac{\mu_\varphi^2}{2} \varphi^2 
              + \frac{\lambda_\varphi}{4}  \varphi^4
	       + \frac{\lambda_4}{2} \varphi^2 H^\dagger H
\label{eq:V1} \\
& 
   + \frac{\mu_1^3}{\sqrt{2}} \varphi 
   + \frac{\mu_3}{2\sqrt{2}} \varphi^3 
   + \frac{\mu}{\sqrt{2}}  \varphi (H^\dagger H) \,,
\label{eq:V2}
\end{align}
where the $\lambda_4$ and $\mu$ terms provide the Higgs portal between the
dark and SM sectors. In order to keep the expressions simple we set in the following
always $\mu_1 = \mu_3 = 0$. Those terms will not introduce new physical
effects and therefore all features relevant for our discussion can be
captured within this restricted framework.

In general mixing between $h$ and $\phi$ will be induced, with physical mass
states $H_1$ and $H_2$ and a mixing angle $\alpha$ with
\begin{equation}
\tan 2\alpha = \frac{\sqrt{2} {\mu v_1} + 2 \lambda_4 v_1v_2}
{2 \lambda_H v_1^2  - 2 \lambda_\phi v_2^2 + \mu v_1^2/(2 \sqrt{2} v_2)} \,.
\end{equation}
%
%
We adopt the convention that for $\alpha \to 0$, $H_1$
corresponds to $h$. Hence, for small mixing and $m_{H_1} = 125$~GeV,
$H_1 \approx h$ becomes a SM-like Higgs.
All direct processes coupling $\chi$ to the SM are proportional to
$\sin^2 2\alpha$ and therefore the mixing angle plays a crucial role
for DM signals.

\subsection{Direct detection}

DM scattering on nuclei relevant for direct detection is mediated via
$t$-channel exchange of the Higgs mass eigenstates $H_1$ and $H_2$. Hence,
scattering is spin-independent. The elastic scattering cross section
$\sigma_{p}$ of $\chi$ off a proton $p$ is obtained as
\begin{equation}\label{eq:DD}
 \sigma_{p}=\frac{g_S^2 \sin^22\alpha}{4 \pi} \, m_{\rm red}^{2} 
 \left(\frac{1}{m_{H_1}^2} - \frac{1}{m_{H_2}^2} \right)^{2} g_{Hp}^{2} \,.
\end{equation} 
The typical size of the scattering cross sections is 
\begin{equation}
\sigma_p \approx 5\times 10^{-43}\,{\rm cm}^2 \, g_S^2 \sin^22\alpha
\left(\frac{m_{\rm red}}{1 \,{\rm GeV}}\right)^2
\left(\frac{1}{m_{H_1}^2} - \frac{1}{m_{H_2}^2} \right)^{2} 
(100 \,{\rm GeV})^{4} \,.
\end{equation}
This number has to be compared to the limit from XENON100, which is 
$\sigma_p \lesssim 10^{-44}\,{\rm cm}^2$ for $m_\chi \simeq
50$~GeV~\cite{Aprile:2011hi}. Hence, couplings of order one and large mixing
are in tension with the bound. In eq.~\eqref{eq:DD} we take into account
only the scalar coupling $g_S$. Similar to the EFT case discussed above, for
pseudo-scalar interactions the cross section is suppressed by $\vel^2\sim
10^{-6}$.

\subsection{LHC Higgs signatures}

In order to define a SM-like Higgs $h$ with $m_{H_1} = 125$~GeV, we will use
the notion of signal strength reduction factor in the event number of a
specific final state of the Standard Model, $X$, in the Higgs boson decay,
see e.g.~\cite{Baek:2011aa, Englert:2011yb, Englert:2011aa}. The latter is
defined as:
\begin{equation}
  r_i\equiv \frac{\sigma_{H_i} {\rm Br}_{H_i\rightarrow X}}
 {\sigma_{H_i}^{\rm SM} {\rm Br}_{H_i\rightarrow X}^{\rm SM}}
\end{equation}
with $i=1,2$ and where $\sigma_{H_i}$ and  
${\rm Br}_{H_i\rightarrow X}$ are the Higgs
production cross section and branching ratio of  $H_i\rightarrow X$,
respectively, while $\sigma_{H_i}^{\rm SM}$ and 
${\rm Br}_{H_i\rightarrow X}^{\rm SM}$ are
the same quantities for a Standard Model Higgs with $m_h=m_{H_i}$. One obtains
\begin{equation}\label{eq:r1r2}
  r_1=\cos^4\alpha \, \frac{\Gamma^{\rm SM}_{H_1}}{\Gamma_{H_1}}  
  \quad \mbox{and} \quad 
  r_2=\sin^4\alpha \, \frac{\Gamma^{\rm SM}_{H_2}}{\Gamma_{H_2}}
\end{equation}
where $\alpha$ denotes the Higgs mixing angle, $\Gamma^{\rm SM}_{H_i}$ is
the total decay width of a SM Higgs of mass $m_h=m_{H_i}$ and
$\Gamma_{H_i}$ is the total decay width of $H_i$ including the decay into
$H_{j \neq i}$ and $\chi$. In order to have a SM-like Higgs we require small
mixing $\alpha$ and identify $H_1$ with the SM Higgs $h$ with $m_{H_1} =
125$~GeV. In practice, we will require that $r_1>0.9$ and $r_2<0.1$. The latter
constraint is imposed to respect the fact that no indication of a second
Higgs-like particle is seen at LHC. In the model under consideration
typically requiring $r_1 > 0.9$ automatically leads to $r_2 < 0.1$. Note
that eq.~\eqref{eq:r1r2} is independent of the Higgs decay channel $X$.
Therefore, we can compare $r_i$ directly with the ATLAS/CMS results on the
signal strength reduction factor obtained from a combination of all search
channels.

\subsection{Numerical results}

We have performed a numerical scan over the parameters of this model using
{\tt micrOMEGAs}~\cite{Belanger:2008sj, Belanger:2010gh}. We assume
$m_{H_1}=125$~GeV and set $g_P = 0$. Then we scan randomly over $m_\chi,
g_S, v_2, \mu, \lambda_4$, and $m_{H_2}$ or $\lambda_\phi$ as free
parameters. In order to ensure perturbativity, we impose that the absolute
value of the couplings $ \lambda_4, \lambda_\phi, \lambda_H$ and $g_S$ are
smaller than $\pi$. For the scalar potential to be bounded from below, we
imposed $\lambda_\phi, \lambda_H>0$ and $\lambda_4>-2
\sqrt{\lambda_\phi\lambda_H}$. We also assume that $\chi$ is the only dark
matter candidate that gives rise to a relic density $0.09<\Omega h^2<0.13$
obtained by thermal freeze-out.  If not mentioned otherwise, we scanned the
following range of parameters: 5~GeV$\leq m_{H_2}, m_\chi\leq 10^4$~GeV,
$10^{-4}\,{\rm GeV}\leq|\mu|, v_2 \leq 10^4$~GeV, and
$10^{-5}\leq|\lambda_4|, |g_S|\leq \pi$. We identify two viable parity conserving 
Higgs portals.

\begin{figure}[tb!]
\begin{center}
\includegraphics[width=10cm]{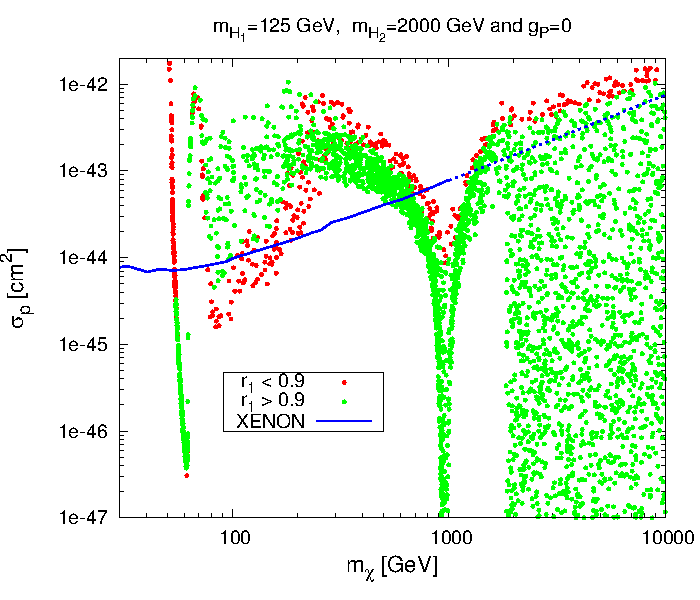}
\mycaption{Proton--DM scattering cross section as a function of the
 dark matter mass in the Higgs portal model for $m_{H_1} = 125$~GeV,
 $m_{H_2} = 2$~TeV, and $g_P=0$. The green points
 correspond to a SM-like $H_1$ with an LHC Higgs signal strength modifier
 $r_1>0.9$, while the red points have $r_1 < 0.9$. 
 The points above the blue line are excluded at 95\%~CL by
 the XENON100 experiment~\cite{Aprile:2011hi}. This exclusion
 limit has been extended for $m_\chi>1$~TeV assuming a linear dependence in $m_\chi$.}
\label{fig:DD2000}
\end{center}
\end{figure}

\subsubsection{Resonant Higgs portal}
We first assume $m_{H_1} \ll m_{H_2}$ and fix $m_{H_2}=2000$~GeV. Requiring
the correct relic abundance we show the predicted direct detection scattering
cross section in fig.~\ref{fig:DD2000} compared to the bound from XENON100.
For DM masses $m_\chi \lesssim 500$~GeV the mediator mass is still ``heavy''
and we recover roughly the EFT behaviour from fig.~\ref{fig:DDEFT}.
However, we clearly observe the suppression of the direct detection rate
when $m_\chi \approx m_{H_{1}}/2$ or $m_{H_{2}}/2$, where there is an
$s$-channel resonance for annihilations, allowing for small coupling
constants while maintaining the correct relic abundance. The red dots in 
fig.~\ref{fig:DD2000} correspond to a signal strength modifier for the Higgs
signal at LHC of $r_1 < 0.9$. Hence, those points would be excluded by
confirming a SM-like Higgs at 125~GeV, while for the green points we have
$r_1 > 0.9$, showing that close to the resonances we can easily have parity
conserving fermionic Higgs portal DM consistent with a SM-like Higgs.

\subsubsection{Indirect Higgs portal}
Let us now discuss the region $m_\chi > m_{H_2} = 2$~TeV in
fig.~\ref{fig:DD2000}. In this case annihilation of $\chi$ into the mediator
becomes kinematically allowed. There are $t$- and $u$-channel diagrams
contributing to this annihilation channel, which are independent of the
mixing angle $\alpha$ and only depend on the coupling $g_S$. Assuming pure
scalar coupling and $m_{H_{1,2}} \ll m_\chi$ we find 
\begin{equation}\label{eq:t-u-chan-annih}
\sigma_{\chi\chi\to \phi\phi} = \frac{3 g_S^4 \vel}{32\pi m_\chi^2} \qquad
\text{($u$- and $t$-channel diagrams)}\,,
\end{equation}
where $\vel$ is the $\chi$ velocity in the CMS. The relic density is
obtained when the reaction $\chi\chi\leftrightarrow \phi\phi$ freezes out.
(Note that for small mixing we have $\phi \sim H_2$.) This fixes essentially
the coupling $g_S$, while leaving the Higgs mixing $\alpha$ unconstrained.
Since the direct detection cross section is proportional to
$\sin^2(2\alpha)$, essentially any value for $\sigma_p$ below the XENON100
bound can be obtained\footnote{At 1-loop DM--nucleus scattering is induced
also for zero Higgs mixing, if $\lambda_4\ne 0$, giving a lower bound on the
scattering cross section. The Wilson coefficient in
eq.~\eqref{eq:scattering} is in this case
$C_1=(\sqrt{2}g_S^2\lambda_4/16\pi^2)(m_\chi v_1/m_\phi^2)f(x)$, with
$x=m_\chi^2/m_\phi^2$ and $f(x)=1/(1-x)-x\log(x)/(1-x)^2$ so that $f(0)=1$.
Note that this means that for zero $\phi-h$ mixing the suppression scale is
$\Lambda_1\sim 16\pi^2 m_\phi^2/m_\chi$ for ${\mathcal O}(1)$ couplings. 
For typical parameter choices the loop process induces tiny cross sections
below $10^{-50}\,\rm cm^2$.}, as confirmed in fig.~\ref{fig:DD2000} for
$m_\chi > 2$~TeV. We study this situation in more detail in the following.  

For $m_{H_{1,2}} < m_\chi$ the exchange of light scalar
fields $H_{1,2}$ between the two annihilating dark matter particles creates
a long range attractive potential (long range compared to the Compton wavelength of $\chi$). As a result there is a {\it Sommerfeld} 
enhancement of the dark matter annihilation
cross-section~\cite{sommerfeld}. This velocity dependent effect has been studied in
detail in several references, see
e.g.~\cite{Hisano:2003ec,ArkaniHamed:2008qn,MarchRussell:2008tu} (see
also~\cite{Arina:2010wv} for a very similar framework).  
In the calculations of the dark matter relic density we estimate the Sommerfeld enhancement
averaged over a thermal distribution at freeze-out temperature $T_f$
following~\cite{Feng:2010zp}. We assume that the cross section determining
the dark matter relic abundance is p-wave suppressed. 
The thermally averaged Sommerfeld factors $ \bar S_\Phi (x_f)$
due to $\Phi=H_1$ and $H_2$ exchanges are
functions of the  couplings $\alpha_{H_1}=
\left(g_S\sin\alpha\right)^2/(4\pi)$ and 
$\alpha_{H_2}= \left( g_S\cos\alpha\right)^2/(4\pi)$, respectively, and of the
dimensionless parameters $\epsilon_\Phi=m_\chi/(\alpha_\Phi m_\Phi)$
and $x_f=m_\chi/T_f$ (all in the notation of~\cite{Feng:2010zp}).  Let us emphasize that in our toy model dark matter does couple to two
mediators, in which case the computation of the
exact Sommerfeld factor is more involved~\cite{McDonald:2012nc} than the results
in~\cite{Feng:2010zp}.  In most of the cases considered here, only one
of the two scalars leads to a non-negligible Sommerfeld correction
$\bar S$ and the relic density is taken to be $\Omega_\chi h^2\propto
1/({\bar S \langle\sigma v\rangle}$), where the thermal averaged
annihilation cross-section $\langle\sigma v\rangle$ is obtained with
{\tt micrOMEGAs}.  If both $H_1$
and $H_2$ lead to a non negligible thermally averaged Sommerfeld factor,
then $\bar S$ is taken to be the largest of the two \footnote{Notice
  that in the particular framework of Ref.~\cite{McDonald:2012nc} it
  was shown that the exchange of multiple mediators can increase the
  Sommerfeld enhancement in the off-resonant region by $\sim 20\%$.
}.  

\begin{figure}[tb!]
\begin{center}
\begin{tabular}{cc}
 \includegraphics[width=7.5cm]{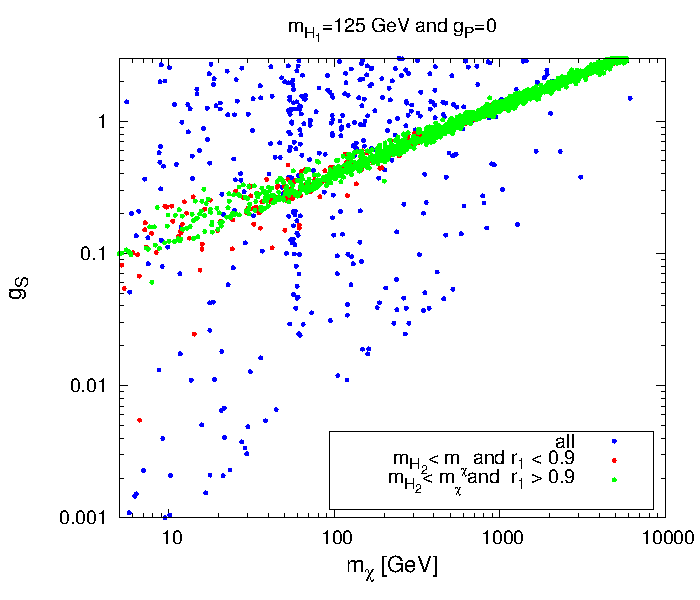} &
 \includegraphics[width=7.5cm]{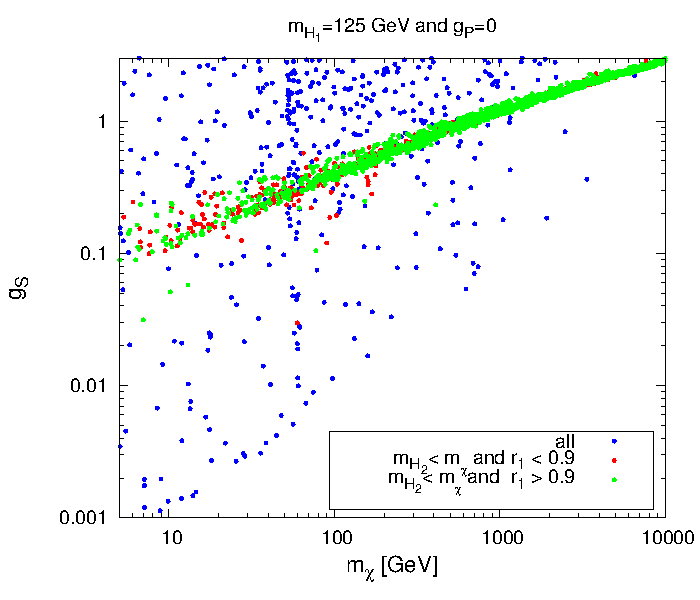}
\end{tabular}
\mycaption{Parameter choices giving rise to a relic density in the WMAP
range in the Higgs portal model with $m_{H_1} = 125$~GeV and $g_P=0$. Green
and red points correspond to $m_{H_2}<m_\chi$ with a more ($r_1>0.9$) or
less ($r_1<0.9$) SM Higgs-like $H_1$, respectively. We show the   scalar coupling $g_S$ as a function of the dark
matter mass without (left) and with (right) Sommerfeld enhancement for the relic density computation. For illustration, we also show the points with $m_{H_2}>m_\chi$
(blue points). } \label{fig:mchi-g}
\end{center}
\end{figure}

For masses $m_\chi\lesssim 100$ GeV, no Sommerfeld enhancement of the
thermal averaged annihilation cross-section is observed. For
$m_\chi\gtrsim 100$ GeV, $\bar S_{H_2}$ can become larger than one and
take values up to 4.5.  Above $m_\chi\sim 1$ TeV, we observe values of
$\bar S_{H_1}\geq 1$ going up to 2. The main impact of the Sommerfeld
enhancement is to allow for smaller values of the couplings $g_S$ at a
given mass $m_\chi$ in order to account for the correct relic
density. This is illustrated in fig.~\ref{fig:mchi-g} where we show
the DM coupling to the scalar singlet, $g_S$, as a function of the DM
mass with and without Sommerfeld enhancement. For the case $m_{H_2} <
m_\chi$ (red and green points) we observe a clear correlation
consistent with $g_S^2 \propto m_\chi$. This is expected when the
relic density is driven by the process $\chi\chi \leftrightarrow
\phi\phi$ according to eq.~\eqref{eq:t-u-chan-annih}. Also notice the
relative flattening of the $g_S^2 -m_\chi$ correlation for $m_\chi$ in
the right panel of fig.~\ref{fig:mchi-g}. This is due to the
presence of the Sommerfeld enhancement factor allowing for smaller
coupling at a given value $m_\chi$ in order to still be consistent
with WMAP data.

\begin{figure}[tb!]
\begin{center}
 \includegraphics[width=8cm]{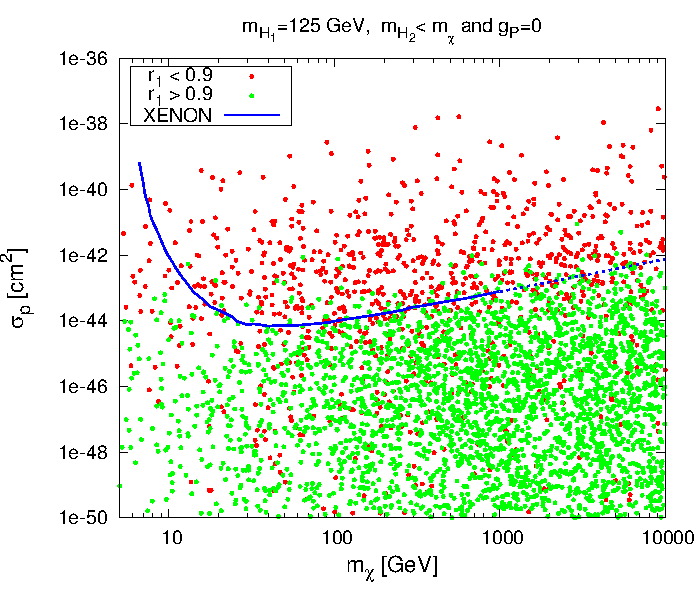} 
\mycaption{Parameter choices giving rise to a relic density in the WMAP
range in the Higgs portal model with $m_{H_1} = 125$~GeV and $g_P=0$. Green
and red points correspond to $m_{H_2}<m_\chi$ with a more ($r_1>0.9$) or
less ($r_1<0.9$) SM Higgs-like $H_1$, respectively. We show the DM--proton
scattering cross section as a function of the dark matter mass for
$m_{H_2}<m_\chi$ only.  The points above the blue line are excluded at 95\%
CL by the XENON100 experiment~\cite{Aprile:2011hi}. This exclusion limit
has been extended for $m_\chi>1$~TeV assuming a linear dependence in
$m_\chi$. \label{fig:mchi-SDDn}}
\end{center}
\end{figure}

In fig.~\ref{fig:mchi-SDDn}, we show the direct detection cross section  requiring
$m_{H_2} < m_\chi$. We see that the $\chi$-nucleon scattering cross section can have vastly varying values. It can be well below the XENON100 bound, while still
accounting for the correct relic density. This confirms that
 the size of the DM annihilation cross section (which controls the relic density)  is no longer related to the strength of DM interactions with the SM. In fig.~\ref{fig:mchi-SDDn}
the green points have $r_1 > 0.9$, which means that $H_1$ will look like a SM Higgs
at the LHC, while the red points have a suppressed Higgs signal, $r_1
< 0.9$. We observe that the $r_1 > 0.9$ requirement tends to keep the
scattering cross section below the XENON100 limit. 

In the relic density calculation we have assumed that the thermal bath of
$\chi$ and $\phi$ has the same temperature as the SM thermal bath. The
contact between those two sectors is provided by the Higgs portal 
$\lambda_4 \varphi^2 (H^\dagger H)$, providing interactions like $\phi\phi
\leftrightarrow hh, \phi \leftrightarrow hh, \phi\phi \leftrightarrow h$. If
those interactions freeze out before $\chi$ decouples from $\phi$, in
principle the dark and visible sectors may acquire different temperatures due
to a change in the number of relativistic degrees of freedom in the visible
sector. Unless both, $\lambda_4$ and $\alpha$, are extremely tiny, this may
change the relic abundance by factors of order one compared to the situation
presented above, while maintaining the qualitative picture. Various
possibilities to obtain the relic abundance for various cases of DM and
mediator properties have been discussed recently in \cite{Chu:2011be}. Note
that as long as the scalar mixing angle $\alpha$ is not exactly zero, $H_2$
is not stable and decays via the Higgs $h$ into SM particles.

Hence, in this situation the Higgs portal acts indirectly, providing the 
link between the dark and visible thermal baths in the early universe. We
call this ``indirect Higgs portal''. This situation is similar to secluded
DM models~\cite{Pospelov:2007mp}, where DM annihilations into light mediator
particles have been discussed, see also \cite{Finkbeiner:2007kk, Kim:2009ke}. A
particular version of the indirect Higgs portal has been obtained in the
model from~\cite{Lindner:2011it}. That model respects a global $U(1)$
symmetry with a complex mediator $\phi$, and the relic density may be set by
the annihilation of DM into the massless Goldstone boson from the
spontaneous breaking of the $U(1)$.

Let us mention that the inclusion of Sommerfeld enhancement in the
computation of the DM relic density does not change qualitatively the
general picture presented here. With respect to indirect detection
searches, notice that the Sommerfeld correction is a velocity
dependent effect that becomes larger when smaller velocities are
involved. The Sommerfeld enhancement that affect dark matter
annihilations in the galactic halo ($v\sim 10^{-6}$) is larger than
$\bar S$ by several orders of magnitude. The annihilation cross
sections involved in the scalar case are however always p-wave
suppressed and we have checked that they for the model under study,
they stay unconstrained by indirect detection searches.

\section{Conclusions}

Motivated by recent hints from LHC experiments for a SM-like Higgs particle
around 125~GeV we have revisited here the possibility that the operator
$(H^\dagger H)$ acts as a portal between the SM and the dark sector. We
adopt the assumption that DM is a fermion, which necessarily requires
additional degrees of freedom to couple it to the Higgs portal. We consider
configurations where those additional particles are heavy and an EFT description
is possible, as well as situations with light mediators. In the latter case
we adopt a simple renormalizable toy model where a real scalar $\phi$ plays
the role of the mediator particle. Assuming further that the DM relic
abundance is obtained by thermal freeze-out in the early universe, the most
simple realization of the fermionic Higgs portal DM is under pressure from
constraints on the DM--nucleon scattering cross section from XENON100. 

We
have identified three simple ways to make thermal fermionic DM consistent
with a SM-like Higgs at 125~GeV and XENON100 bounds:
\begin{itemize}
\item
{\it Pseudo-scalar Higgs portal.} If DM couples to the Higgs portal via
$\bar\chi \gamma_5 \chi$ the direct detection cross section is suppressed by
the DM velocity $\vel^2 \sim 10^{-6}$, whereas the annihilation cross
section responsible for the relic abundance is unsuppressed.
\item
{\it Resonant Higgs portal.} If the DM mass $m_\chi$ is close to half of the
Higgs mass $m_h$ or the mediator mass $m_\phi$, then the annihilation cross
section is enhanced by an $s$-channel resonance, allowing for small
couplings and a suppressed direct detection cross section.
\item
{\it Indirect Higgs portal.} If the mediator $\phi$ is lighter than the DM
$\chi$, the relic density can be obtained by $\chi\chi \leftrightarrow
\phi\phi$ annihilations, where the $t$- and $u$-channel diagrams are
independent of the Higgs portal strength. The Higgs portal only acts
indirectly to provide thermal contact between the dark and the visible
sector thermal baths.
\end{itemize}

In all cases it is possible to have a SM-like Higgs, with an LHC signal
strength modifier $r_1 > 0.9$ (where $r_1 = 1$ corresponds to the SM Higgs).
This framework is sometimes called ``LHC nightmare scenario'', with no
new-physics signal at LHC apart from a SM-like Higgs. Also, by construction,
the models discussed here can have unobservably small signals in direct
detection experiments. However, in general a signal can be expected for
indirect detection. For the pseudo-scalar and the indirect Higgs portals we
predict a conventional indirect detection signal (dominated by annihilations
into $\bar b b$ or gauge bosons), with an annihilation cross section
determined by the thermal freeze-out of $\sigma \vel \simeq 3\times
10^{-26}\,\rm cm^3s^{-1}$. In the case of resonant Higgs portal there might
be also an enhancement of the annihilation cross section today compared to
the one in the early universe \cite{Ibe:2008ye,Feldman:2008xs} if the resonance is combined
with a pseudo-scalar coupling. However, the enhancement effect may be not
enough to overcome the velocity suppressed annihilation rate for pure scalar
couplings. 

In conclusion, fermionic Higgs portal DM remains a viable option if a
SM-like Higgs should be established at the currently hinted mass of 
around 125~GeV. We have outlined simple mechanisms to obtain a classic
``WIMP'' DM candidate, whose relic abundance is set by thermal freeze-out,
with no DM related signal at the LHC and highly suppressed rates in direct
detection experiments, but still potentially observable in indirect detection.

\bigskip
{\bf Acknowledgement.} We thank Joachim Kopp and Yasutaka Takanishi for
useful discussions. L.L.H and T.S. acknowledge partial support from the  European Union FP7  ITN 
INVISIBLES (Marie Curie Actions, PITN- GA-2011- 289442).

\bibliographystyle{my-h-physrev.bst}
\bibliography{./refs}

\end{document}